# Dielectric signature of charge order in lanthanum nickelates

P. Sippel[1], S. Krohns[1], E. Thoms[1], E. Ruff[1], S. Riegg[1], H. Kirchhain[1], F. Schrettle[1], A. Reller[2], P. Lunkenheimer[1,*], and A. Loidl[1]

[1]Experimental Physics V, Center for Electronic Correlations and Magnetism, University of Augsburg, 86159 Augsburg, Germany
[2]Department for Resource Strategy, University of Augsburg, 86159 Augsburg, Germany

**Abstract.** Three charge-ordering lanthanum nickelates $La_{2-x}A_xNiO_4$, substituted with specific amounts of $A$ = Sr, Ca, and Ba to achieve commensurate charge order, are investigated using broadband dielectric spectroscopy up to GHz frequencies. The transition temperatures of the samples are characterized by additional specific heat and magnetic susceptibility measurements. We find colossal magnitudes of the dielectric constant for all three compounds and strong relaxation features, which partly are of Maxwell-Wagner type arising from electrode polarization. Quite unexpectedly, the temperature-dependent colossal dielectric constants of these materials exhibit distinct anomalies at the charge-order transitions.

## 1 Introduction

Electron correlations in condensed matter have been in the focus of interest in recent years as they give rise to a rich variety of phenomena like high-$T_c$ superconductivity, metal-insulator transitions, colossal magnetoresistance, and multiferroicity [1,2,3,4]. Charge order (CO) is a typical and intensely investigated example for strong electron correlations leading to the ordering of electrons in patterns like checkerboards or stripes at temperatures $T < T_{CO}$ [1,5,6,7]. This usually is accompanied by a significant increase of resistivity or even a metal-insulator transition when the formerly delocalized electrons condense into the charge-ordered phase. While dielectric spectroscopy does not belong to the standard methods to investigate CO, a number of dielectric studies have revealed interesting dielectric properties of these materials like so-called colossal dielectric constants and anomalies at the CO transition like peaks or jumps in the temperature-dependent dielectric constant $\varepsilon'(T)$ [8,9,10,11,12,13,14,15,16]. Even electronic ferroelectricity originating from CO has been considered [17,18,19,20]. While very high values of $\varepsilon'$ often can be ascribed to non-intrinsic electrode or grain-boundary effects [21,22,23], which indeed has been explicitly demonstrated for some charge ordered materials [16,24], it seems well possible that at least some of the published dielectric phenomena are connected to the CO [10,16].

A prototypical and often investigated charge ordering system is $La_{2-x}Sr_xNiO_4$, which exhibits CO in large parts of its phase diagram [25,26,27,28]. Strontium substitution on the La sites of $La_2NiO_4$ introduces holes at the Ni sites (i.e., $Ni^{3+}$ instead of $Ni^{2+}$), which at $T < T_{CO}$ no longer are statistically distributed but order in stripes within the $NiO_2$ planes of the crystal structure [5,26,27]. In addition, at a temperature $T_{SO} < T_{CO}$ a spin order transition was found, which corresponds to an antiferromagnetic ordering within the stripes formed by the $Ni^{2+}$ ions [27,28]. Charge-stripe order was also detected in the relatively rarely investigated Ca- and Ba-substituted compounds [7,29,30]. We have shown earlier that the dielectric constant of $La_{15/8}Sr_{1/8}NiO_4$ reaches values as high as $10^6$ and that this compound retains its colossal values of the dielectric constant up to the GHz frequency range making it a promising material for capacitor applications [16,31]. However, no trace of the CO transition was found in the temperature-dependent dielectric properties. In fact, $T_{CO}$ is not known for $x = 1/8$ as the lowest Sr concentration, for which it was unequivocally determined, is 0.135 [26].

CO can be expected to be best pronounced for commensurate values of $x$ close to 1/3 or 1/2 [7,32]. In the present work, we provide a dielectric characterization of $La_{1.7}Sr_{0.3}NiO_4$ with $T_{CO} \approx 220$ K [28,33]. In addition, we have investigated Ba- and Ca-substituted $La_2NiO_4$, both for $x = 0.33$. Both systems are known to show CO transitions at similar temperatures as the strontium system [7,29,30], which here is corroborated by specific-heat and magnetic-susceptibility experiments. Temperature-dependent dielectric measurements of all three sample materials are performed at various frequencies extending up to about 2 GHz. This allows obtaining information on both the bulk dielectric properties, usually revealed at high frequencies only, and the heterogeneity-related colossal $\varepsilon'$ values, seen at the lower frequencies [16,23]. This work has been motivated by the idea that stripe order creates internal microscopic interfaces between metallic and insulating layers, hence establishing a self-organized multilayer capacitor with large dielectric constant. Indeed, we find well-pronounced anomalies in the dielectric properties at the CO transition in all three materials.

## 2 Experimental details and sample preparation

Polycrystalline samples were prepared via classical solid-state reactions, using stoichiometric mixtures of the binary oxides $La_2O_3$ (previously dried 6 hours at 800°C), NiO and the alkaline earth carbonates. The powders were well ground in agate mortars and calcined 48 hours at 1100 °C in air with one intermediate grinding after 24 hours. For the heat-capacity measurements, approximately 20 mg of the sample powders were pressed into pellets of 3 mm diameter. The pellets were sintered in highly densified aluminum oxide crucibles for 48 hours at 1050 °C in air. Room-temperature x-ray diffraction with subsequent Rietveld analysis documents that all $La_{2-x}A_xNiO_4$ compounds investigated exhibit the well-known tetragonal structure with the proper I4/mmm space group (no. 139). No impurity phases were detected

---

[a] e-mail: peter.lunkenheimer@physik.uni-augsburg.de

above experimental uncertainty. The lattice constants have been determined as follows: $A = Sr_{0.3}$: $a = 3.831$ Å, $c = 12.707$ Å; $A = Ca_{0.33}$: $a = 3.822$ Å, $c = 12.594$ Å; $A = Ba_{0.33}$: $a = 3.848$ Å, $c = 12.830$ Å and are close to values reported in literature (see, e.g., [29,30])

The specific heat was measured between 2 and 300 K using a Quantum Design physical-properties measurement system (PPMS). For the dielectric measurements, silver paint or sputtered silver contacts were applied at opposite sides of the plate-like samples. The dielectric properties at frequencies from 20 Hz to 1 MHz were determined using a frequency-response analyzer (Novocontrol alpha) [34]. Measurements between 1 MHz and 3 GHz were performed with a coaxial reflection technique, employing impedance analyzers (Hewlett-Packard 4291B and Agilent E4991A) [34,35]. Applied ac voltages in these experiments were 1 V (low frequencies) and 0.4 V (high frequencies). For sample cooling, a closed-cycle refrigerator and a helium bath cryostat were used. The dc-conductivity was measured using a standard 4-point setup; here the sample was cooled using the PPMS.

## 3 Experimental results and discussion

The specific heat of $La_{2-x}Sr_xNiO_4$ with $x \approx 1/3$ is known to exhibit a well-pronounced peak at the CO transition but no anomaly at the spin-order transition [36]. Figure 1 shows the specific heat $C_p$ of the Ca- and Ba-substituted systems. The behavior of $La_{1.67}Ba_{0.33}NiO_4$ strongly resembles the one found in the corresponding Sr-substituted compound [36]: There is a pronounced peak that can be ascribed to the CO transition ($T_{CO} = 226 (\pm 0.5)$ K) but no signature of the spin-order transition is observed. In the Ca-compound, a weaker anomaly at roughly the same temperature ($T_{CO} = 227 (\pm 3)$ K) shows up. The absence of an anomaly in $C_p(T)$, corresponding to the spin order in $La_{1.67}Sr_{0.33}NiO_4$, was attributed to the 2-dimensional nature of the magnetic order in the nickelates [36]. The found CO-transition temperatures of 226 K and 227 K are in reasonable accord with those reported in Ref. [29], where values of 224.3 K (Ca) and 224.9 K (Ba) were obtained from resistivity results, and with $T_{CO} = 225$ K deduced in Ref. [7] from resistivity and susceptibility measurements. For $La_{1.67}Sr_{0.33}NiO_4$, values of $T_{CO}$ between 235 K and 239 K were reported [7,29,36].

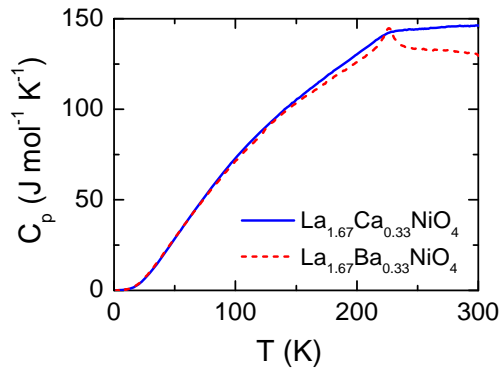

**Fig. 1.** Temperature-dependent specific heat $C_p$ of the Ca- and Ba-substituted compounds.

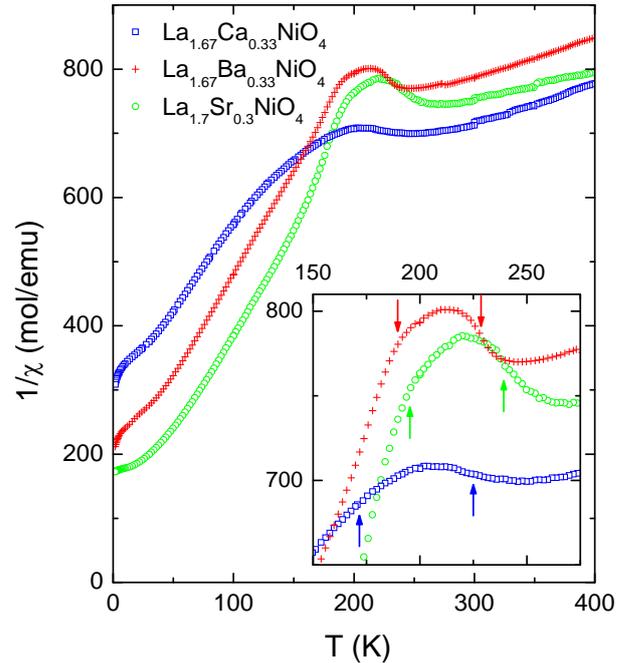

**Fig. 2.** Temperature-dependent inverse magnetic susceptibility $1/\chi$ of the investigated compounds, measured at an external magnetic field of 0.1 T. The inset shows a magnified view of the region of the phase transitions. The right arrows indicate $T_{CO}$ (listed in Table 1). The left arrows point to anomalies ascribed to the spin-order transitions at $T_{SO}$ (Table 1).

In Fig. 2 the temperature dependence of the magnetic susceptibility $\chi$, measured at an external magnetic field of 1 T, is provided for a further characterization of the investigated samples. The findings are similar to those of Cheong *et al.* [7]. The plot of $1/\chi$ signifies that, at best, $\chi(T)$ does follow a Curie- or Curie-Weiss law in very limited temperature ranges only. Moreover, especially at high temperatures unreasonable parameters would result from such a description. In agreement with previous reports [7,36], the CO transition shows up as a steplike behavior (increase of $1/\chi$ with decreasing temperature). The estimated temperatures of the midpoints of these anomalies (right arrows in the inset of Fig. 2; for the temperatures, see Table 1) reasonably agree with the present results from specific heat (Fig. 1) and from literature [7,36]. It is well known that the spin order only leads to a weak anomaly in $\chi(T)$ [36], which is indicated by the left arrows in the inset of Fig. 2 (for the temperatures, see Table 1). The anomaly in the magnetic susceptibility corresponds to a somewhat stronger decrease of $1/\chi$ (increase of $\chi$) on decreasing temperatures. The increase in the susceptibility assigned to $T_{SO}$ signals the onset of magnetic order and the concomitant formation of a weak spontaneous ferromagnetic moment probably driven by spin canting [36]. For the Ca-compound, this anomaly is hardly visible and the obtained temperature is a rough estimate only. For the Sr-system, $T_{SO}$ reasonably agrees with the values of 194 K reported in [36]. For the Ca- and Ba-substituted materials, to our knowledge $T_{SO}$ has not been previously reported.

**Table 1.** Transition temperatures (in K) of the three investigated samples, derived from Figs. 1 and 2.

|    | $T_{CO}$ (from $C_p$) | $T_{CO}$ (from $\chi$) | $T_{SO}$ (from $\chi$) |
|----|----|----|----|
| Sr | -         | 239 ($\pm$2)   | 195 ($\pm$2)   |
| Ca | 227 ($\pm$3)   | 225 ($\pm$3)   | 172 ($\pm$10)  |
| Ba | 226 ($\pm$0.5) | 228 ($\pm$2)   | 190 ($\pm$2)   |

Figure 3 shows the temperature-dependent dielectric constant, conductivity, and resistivity of $La_{1.7}Sr_{0.3}NiO_4$ for a variety of measurement frequencies, $\nu > 1$ MHz. Similar to the previously investigated single-crystalline compound with $x = 1/8$ [16], colossal values of the dielectric constant ($\varepsilon' \approx 10000$) are observed at the higher temperatures and lower frequencies (Fig. 3(a)). In addition, there is the typical steplike decrease of $\varepsilon'(T)$ with decreasing temperature, whose onset strongly shifts to lower temperatures with decreasing frequency. It can be ascribed to a Maxwell-Wagner relaxation, just as the qualitatively similar behavior reported for $x = 1/8$ in Ref. [16]. As treated in detail, e.g., in [23], Maxwell-Wagner relaxations arise from dielectric heterogeneities in the sample, i.e. the presence of two (or more) regions with different dielectric properties. If one of the regions is of interfacial character (i.e., thin and highly resistive), it introduces a large capacitance into the system, dominating the dielectric behavior at low frequencies. At high frequencies and/or low temperatures, this capacitor is shorted and the intrinsic behavior is observed [22,23].

In Ref. [16] we have speculated that the internal heterogeneities introduced by the charge order at least partly contribute to the high dielectric constants found in $La_{15/8}Sr_{1/8}NiO_4$. However, we also have demonstrated that contact effects provide by far the strongest contribution to the colossal dielectric constants in this compound. In this context, it is an intriguing finding that obviously the colossal dielectric constant detected at the upper plateau in the $\varepsilon'(T)$ curves of Fig. 3(a) shows a clear anomaly at around 235 K, close to the CO transition, with $\varepsilon'$ decreasing by about 5000 when crossing $T_{CO}$. This plateau corresponds to the limiting values of $\varepsilon'(\nu)$ for low frequencies, i.e. the static dielectric constant $\varepsilon_s$ of the observed relaxation, whose temperature dependence is indicated by the solid line in Fig. 3(a).

Another, at first glance even more spectacular anomaly is revealed at the highest frequencies in Fig. 3(a): $\varepsilon'(T)$ bends down when the temperature approaches $T_{CO}$ from below and even becomes negative (the negative values do not show up in the semi-logarithmic representation of Fig. 3(a)). However, this behavior can be simply ascribed to the significant increase of the dc conductivity (solid line in (b)), which occurs when the charge carriers successively become released from the ordered structure when approaching $T_{CO}$ [29]. Then the inductive properties of the sample start to dominate its total impedance and $\varepsilon'$ no longer is detectable but appears to be negative [37,38]. This occurs especially at high frequencies, where the impedance $Z_L$ arising from the inductance $L$ of the sample becomes large ($Z_L = i2\pi\nu L$). Quite generally, $\varepsilon'$ and $\sigma'$ are not well-suited quantities for highly conducting samples [39]. It is notable, that, in contrast to $La_{15/8}Sr_{1/8}NiO_4$ [16], for $x = 0.3$ $\varepsilon'$ does not retain its colossal magnitude well up to the GHz range at room temperature. However, it at least remains colossal up to about 100 MHz, still better than the prominent colossal-dielectric-constant material $CaCu_3Ti_4O_{12}$ [40,41]. Nevertheless, the dielectric loss in this highly doped material (not shown) clearly is too high for any capacitor applications.

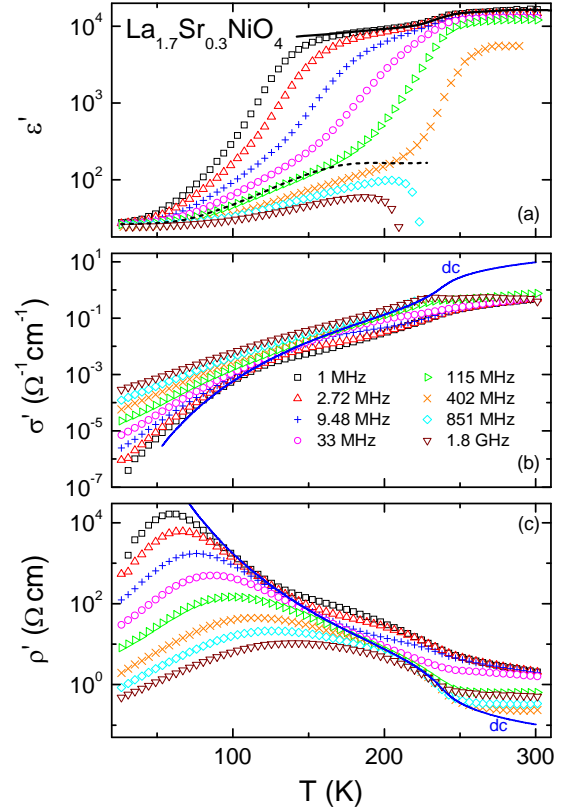

**Fig. 3.** Temperature dependence of the dielectric constant (a), conductivity (b), and resistivity (c) of $La_{1.7}Sr_{0.3}NiO_4$ shown for selected measurement frequencies. The solid line in (a) indicates the temperature dependence of the static dielectric constant. The dashed line illustrates the possible presence of a second relaxation step at low temperatures. The line in (c) is the resistivity from a four-point dc-measurement, scaled to match the present data in the region of about 100 - 130 K. The line in (b) corresponds to the same data in the conductivity representation.

The measured ac conductivity presented in Fig. 3(b) shows a change of slope at $T_{CO}$ for the higher frequencies but no increase at $T > T_{CO}$, as expected from the dc conductivity, is detected. This apparent discrepancy is made obvious by comparing the ac results to the measured dc conductivity (line in Fig. 3(b) [42]). It can be ascribed to effects of the inductance, too. Inductivity contributions can be modeled by an inductance connected in series to the sample. This inductance contributes to both the real and imaginary part of the admittance $Y = 1/Z$ of the sample and thus also to $\varepsilon'$ and $\sigma'$. However, the real part of the impedance $Z$ (i.e., the resistance) should remain unaffected by the inductance, which contributes to the imaginary part of $Z$ only [39]. Thus, for highly conducting samples, plotting the ac resistivity $\rho'$ often proves useful [37,38,43]. In Fig. 3(c), $\rho'$ of $La_{1.7}Sr_{0.3}NiO_4$ is provided. For an example of a detailed interpretation of such type of plots, the reader is referred to Ref. [38]. The dc resistivity $\rho_{dc}$ can be estimated from this plot by the regions of weak frequency dependence. For temperatures approaching the CO transition, the estimated dc resistivity agrees reasonably with the results from the four-point dc-measurement (line) and they also are in accord with the dc-results reported by Han *et al.* [29] (not shown). This especially includes the onset of the decrease of $\rho_{dc}$ when approaching $T_{CO}$ from low temperatures. However, beyond $T_{CO}$, $\rho'(T)$ levels off and for all frequencies remains larger than the further decreasing dc resistivity. These deviations can be explained by the skin effect, i.e. the resistance is enhanced in the ac measurements due to the fact that the current becomes restricted to the outer regions of the sample at high frequencies [38,43]. Generally the skin effect becomes important for highly conducting samples, which here is the case at $T > T_{CO}$. In Ref. [38] an example for an exact evaluation of such behavior using an equivalent-circuit analysis is provided but this is out of the focus of the present investigation. Overall, the most significant finding of Fig. 3, not explainable by inductance or skin-effect contributions, is the anomaly in $\varepsilon'(T)$ at low frequencies, $\nu < 10 - 100$ MHz, close to $T_{CO}$.

The continuous increase of $\sigma'$ with increasing frequency, exceeding $\sigma_{dc}$, which is observed at low temperatures ($T < 150$ K) in Fig. 3(b), is due to hopping conductivity [37,38,43]. Hopping charge transport in this compound is in accord with the reported finding of $T^{-1/4}$ temperature dependence of $\log(\rho_{dc})$ [29], characteristic of variable range hopping. The decrease of $\rho'(\nu)$ with increasing frequency at low temperatures (Fig. 3(c)) partly can be ascribed to this phenomenon, too. In addition, the shortening of the sample capacitance at high frequencies contributes here, finally leading to a peak in $\rho'(T)$. This mirrors the fact that in this high-resistivity region $\rho'$ is quite unsuited to provide information on the intrinsic sample behavior [39].

Figure 4 reveals that the dielectric constants of $La_{1.67}Ca_{0.33}NiO_4$ and $La_{1.67}Ba_{0.33}NiO_4$ exhibit a qualitatively similar behavior as for Sr-substitution. Again, in $\varepsilon'(T)$ the typical Maxwell-Wagner relaxation steps and the inductance-induced approach of negative values at $T > T_{CO}$ for the highest frequencies are observed. Most importantly, for both materials $\varepsilon_s(T)$ (i.e., the upper plateau of the relaxation steps, indicated by the solid lines) shows a clear anomaly close to the CO temperatures determined from specific heat and magnetic susceptibility (Table 1). For $La_{1.67}Ca_{0.33}NiO_4$ it was necessary to perform additional measurements at $\nu < 1$ MHz in order to observe this anomaly, unobstructed from the onset of the relaxation step. The latter results reveal an unusual further continuous decrease of $\varepsilon'$ at temperatures below the anomaly at $T_{CO}$. Similar behavior can also be suspected for the other compounds. Currently we have no explanation for this phenomenon.

Interestingly, in the Ca-substituted material clear evidence for a second relaxation step of lower amplitude is found, occurring at lower temperatures than the main relaxation (exemplified for the 1 MHz curve by the dashed line in Fig. 4(a)). A closer inspection of the results for the Sr- and Ba-compounds reveals similar indications of such a second relaxation showing up as a weak shoulder below the main $\varepsilon'(T)$ step (dashed lines in Figs. 3(a) and 4(b)). In Ref. [16] some evidence for such a second relaxation was also reported for $La_{15/8}Sr_{1/8}NiO_4$. There it was speculated that this second step might be of "quasi-intrinsic" nature and due to a Maxwell-Wagner relaxation that arises from the internal heterogeneities caused by the charge ordering.

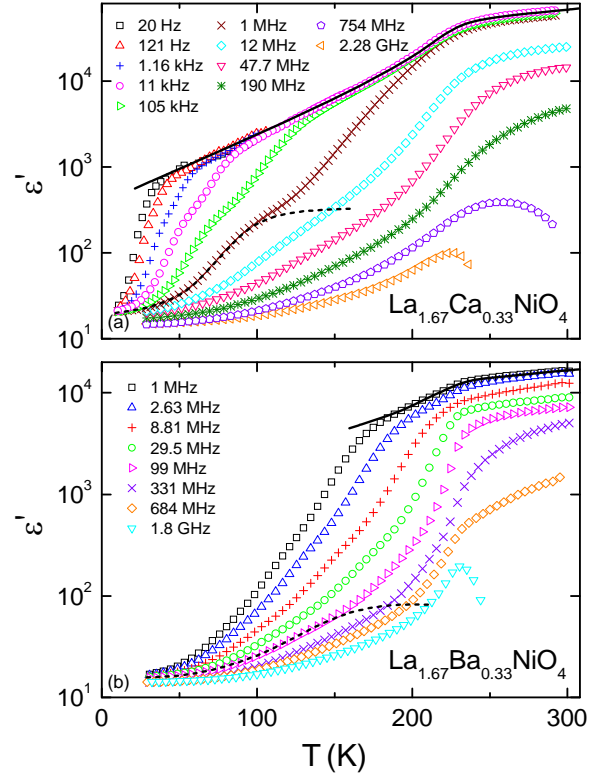

**Fig. 4.** Temperature dependence of the dielectric constant of $La_{1.67}Ca_{0.33}NiO_4$ (a) and $La_{1.67}Ba_{0.33}NiO_4$ (b) for selected measurement frequencies. The solid lines indicate the temperature dependence of the static dielectric constant. The dashed lines demonstrate the likely presence of a second relaxation step at low temperatures.

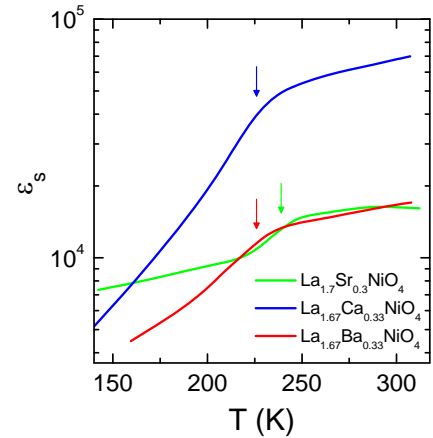

**Fig. 5.** Temperature dependence of the static dielectric constant as estimated from Figs. 3(a), 4(a) and (b). The arrows indicate the CO transition temperatures 227 K (Ca), 226 K (Ba), and 239 K (Sr); see Tab. 1.

Fig. 5 shows the static dielectric constant of the three materials as estimated from Figs. 3(a) and 4. The mentioned anomalies nicely show up as changes of slope in $\varepsilon_s(T)$, close to the respective charge-order temperatures of the three materials, which are indicated by the arrows in Fig. 5. Obviously, for the Sr-system it occurs at a slightly higher temperature than for the Ca- and Ba-compounds, fully consistent with the higher $T_{CO}$ for this material (Table 1). A close inspection of Fig. 5 reveals that in all cases the observed change of slope sets in at somewhat higher temperature than $T_{CO}$, which can be ascribed to the fact that CO fluctuations already exist at temperatures significantly above $T_{CO}$. A similar behavior is also found for the dc conductivity (line in Fig. 3(b)).

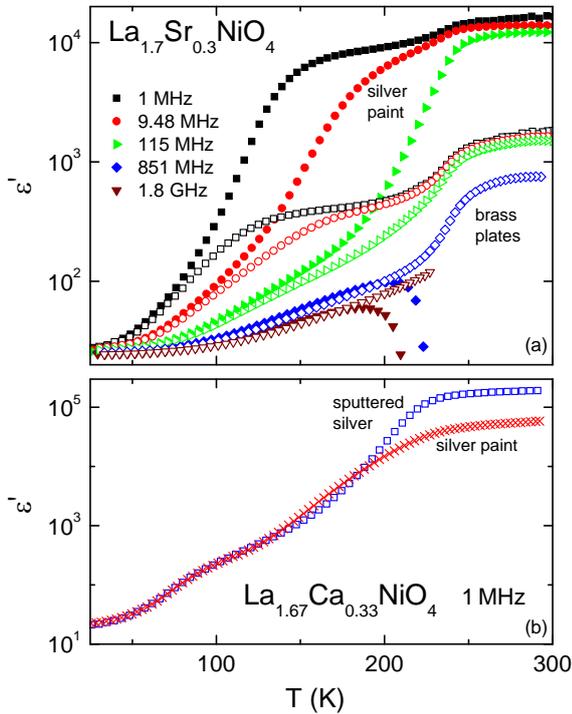

**Fig. 6.** (a) Temperature dependence of the dielectric constant of $La_{1.7}Sr_{0.3}NiO_4$, shown for selected measurement frequencies. The closed and open symbols denote results obtained with silver paint or brass-plate contacts, respectively. (b) $\varepsilon'(T)$ of $La_{1.67}Ca_{0.33}NiO_4$ at 1 MHz, measured with sputtered silver and silver-paint contacts.

As the CO transition leads to a clear anomaly in the static dielectric constant in all three materials, the question arises if the observed relaxation and colossal values of $\varepsilon'$ are directly caused by the CO. To check this conjecture, just as for $La_{15/8}Sr_{1/8}NiO_4$ [16] we have performed measurements with different contact types. Figure 6(a) shows the results for the Sr-compound. Here the closed symbols represent curves that already were shown in Fig. 3(a), measured with silver-paint contacts. The data represented by the open symbols were obtained by putting the same sample between two polished brass plates after removing the silver paint. While the anomaly at the CO transition is well visible in both cases, obviously the variation of the contact type has a strong influence on the absolute values of $\varepsilon'$ at low frequencies, i.e. $\varepsilon_s$. Similar results were also obtained for the Ca-substituted compound, for which measurements with silver paint and sputtered contacts were performed (Fig. 6(b)). These findings strongly point to a contact-related origin of the colossal dielectric constants. It can be explained when considering the different "wetting" of the contact surface with the metal for the different contact types. This should have a dramatic effect on the formation of Schottky diodes that are suspected to form at the electrode-sample interface of these semiconducting samples [16,22,23].

At first glance, the results of Fig. 6 seem confusing: If the high $\varepsilon'$ values are dominated by non-intrinsic electrode effects, why does the intrinsic CO transition have such strong influence on the measured results in this contact-dominated region? Based on the current data, a definite answer to this question is difficult. One could speculate that the observed variation of $\varepsilon_s$ at $T_{CO}$ is due to a contribution from the above-mentioned second relaxation with weaker amplitude. This relaxation does not seem to change when the contact material is varied (Fig. 6) and may arise from internal heterogeneities caused by the charge ordering [16]. Then, of course, it should be strongly affected by the CO transition. However, this scenario seems unlikely because the static dielectric constant of this relaxation can be estimated to be of the order of several 100 (see dashed lines in 3(a) and 4) and it thus is implausible that it should provide any significant contribution to the overall $\varepsilon_s$, which is of the order of $10^4$.

Alternatively, the following scenario should be considered: It seems reasonable that the depletion layer forming at the sample-electrode interface should be influenced by the intrinsic properties. When they change at $T_{CO}$, its layer thickness may change, too, and cause the observed variation of $\varepsilon_s$. In case of the generation of a Schottky-diode at the metal-semiconductor interface, the thickness of the insulating layer is given by

$$d = \sqrt{\frac{2\varepsilon\varepsilon_0\Phi}{eN}}, \qquad (1)$$

where $e$ is the electron charge, $\Phi = (\varphi_m - \varphi_s)/e$ is the difference of the electron work functions of the metal and semiconductor, $N$ is the charge carrier concentration, $\varepsilon$ is the intrinsic dielectric constant of the semiconductor, and $\varepsilon_0$ is the permittivity of vacuum. Three quantities in this expression, namely $\varepsilon$, $\Phi$, and $N$, in principle may be influenced by the CO transition and induce a change of the depletion-layer thickness (and thus of the apparent $\varepsilon'$). A variation of the number of free charge carriers seems to be the most plausible effect: $N$ is expected to reduce below $T_{CO}$ because previously free charge carriers no longer participate in the charge transport when they become locked into fixed positions in the charge-ordered state. This fact also is mirrored by the observed drop of $\sigma_{dc}$ (Fig. 3(b)). This reduction of $N$ should lead to an increase of $d$ (Eq. (1)) and thus may explain the observed decrease of $\varepsilon_s$ (because $\varepsilon_s$ is proportional to the layer capacitance $C$ and $C \propto 1/d$). Within this scenario, the variation of $\varepsilon_s$ should be proportional to the square root of the conductivity variation. However, this relation only holds approximately: For $La_{1.7}Sr_{0.3}NiO_4$ the dc conductivity varies by about a factor of four (cf. line shown in Fig. 3(b)) while in Fig. 6(a) the $\varepsilon_S$ variation is roughly 1.8 for silver-paint and 2.6 for brass-plate contacts. Of course, a number of oversimplifying assumptions were made in the above model, e.g., that the $\sigma_{dc}$ variation is solely due to a variation of $N$ and not of the charge-carrier mobility and that $\varepsilon$ and $\Phi$ remain constant at the transition.

## 4 Summary and conclusions

In summary, we have investigated the temperature and frequency-dependent dielectric properties of three charge-ordering lanthanum nickelates, with doping levels close to 1/3, favoring the formation of commensurate CO. For the rarely investigated Ca- and Ba-substituted compounds we find nearly identical CO transition temperatures of about 226-227 K (Table 1) and spin-order temperatures of 172 K (Ca) and 190 K (Ba). All three compounds exhibit colossal magnitudes of the dielectric constant of the order of $10^4$. Just as for $La_{15/8}Sr_{1/8}NiO_4$, two superimposed relaxation processes are found, one of them being clearly related to electrode effects. Close to the CO transition, the three materials exhibit marked anomalies in the temperature dependence of their colossal dielectric constants. Obviously, this phenomenon mirrors the variation of the intrinsic material properties at the transition, most likely affecting the formation of the depletion layer at the electrode-sample interface. However, the details of the underlying mechanism are not clarified yet.

Overall, our results show that the colossal dielectric constants observed in transition-metal oxides, despite mostly of non-intrinsic origin, can be strongly influenced by changes of the intrinsic sample properties. The details of this behavior and of other properties of colossal dielectric constants in general (e.g., the further decrease of $\varepsilon_s$ for $T \rightarrow 0$ observed in the present work) are not well understood yet and their further investigation should help enhancing our knowledge of the intrinsic properties of the involved materials.

This work was supported by the Deutsche Forschungsgemeinschaft via the Transregional Collaborative Research Center TRR80 (Augsburg, Munich).